\documentclass[%
 aip,
 amsmath,amssymb,
 reprint,%
]{revtex4-1}

\usepackage{graphicx}
\usepackage{dcolumn}
\usepackage{bm}

\usepackage[utf8]{inputenc}
\usepackage[T1]{fontenc}
\usepackage{mathptmx}
\usepackage{etoolbox}
\usepackage{siunitx} 
\usepackage{textcomp}

\makeatother
\begin{document}

\preprint{AIP/123-QED}

\title[Enhanced light collection from a gallium nitride color center using a near index-matched solid immersion lens]{Enhanced light collection from a gallium nitride color center using a near index-matched solid immersion lens}
\author{S.G. Bishop}
 \affiliation{School of Engineering, Cardiff University, Queen's Building, The Parade, Cardiff, UK, CF24 3AA}
\author{J.P. Hadden}
\affiliation{School of Physics and Astronomy, Cardiff University, Queen's Building, The Parade, Cardiff, UK, CF24 3AA} 
\author{R. Hekmati}
\affiliation{School of Physics and Astronomy, Cardiff University, Queen's Building, The Parade, Cardiff, UK, CF24 3AA}
\author{J.K. Cannon}
 \affiliation{School of Engineering, Cardiff University, Queen's Building, The Parade, Cardiff, UK, CF24 3AA}

\author{W.W. Langbein}
\affiliation{School of Physics and Astronomy, Cardiff University, Queen's Building, The Parade, Cardiff, UK, CF24 3AA}
\author{A.J. Bennett}
 \affiliation{School of Engineering, Cardiff University, Queen's Building, The Parade, Cardiff, UK, CF24 3AA}
 \email{BennettA19@cardiff.ac.uk}

\begin{abstract} 
Among the wide-bandgap compound semiconductors, gallium nitride is the most widely available material due to its prevalence in the solid state lighting and high-speed/high-power electronics industries. It is now known that GaN is one of only a handful of materials to host color centers that emit quantum light at room temperature. In this paper, we report on a bright color center in a semi-polar gallium nitride substrate, emitting at room temperature in the near-infrared. We show that a hemispherical solid immersion lens, near index matched to the semiconductor, can be used to enhance the photon collection efficiency by a factor of $\SI{4.3\pm0.1}{}$, whilst improving the lateral resolution by a factor equal to the refractive index of the lens.
\end{abstract}

\maketitle

Color centers in semiconductor materials are promising sources of non-classical photon states. Owing to their localised energy levels, which are embedded deep within the bulk electronic band gaps of their host, they combine the optical properties of single atoms with the scalability of a solid-state environment. Indeed, enhanced electronic confinement arising from mid-gap energy levels in wide-bandgap semiconductors enables single photon emission at room temperature and above\cite{Fu2020}. A number of such color centers have been discovered, including in diamond\cite{Fu2020,Kurtsiefer2000,Wang2006,Maze2008,Dutt2007,Balasubramanian2009,Hadden2010,Knowles2014}, silicon carbide (SiC)\cite{Koehl2011,Widmann2015}, aluminium nitride (AlN) \cite{Bishop2020}, gallium nitride (GaN)\cite{Berhane2017,Zhou2018,Nguyen2019} and hexagonal-boron nitride (h-BN)\cite{Tran2016}. Discovery of these color centers has lead to impressive demonstrations of quantum technologies, including; nanoscale magnetic sensing\cite{Maze2008}, nanoscale quantum thermometry\cite{Delord2020}, quantum repeaters\cite{Bhaskar2020}, coherent control of isolated quantum spin states\cite{Dutt2007,Balasubramanian2009,Koehl2011,Knowles2014,Widmann2015}, a room temperature quantum light emitting diode\cite{Lohrmann2015} and the first room temperature continuous maser\cite{Breeze2018}. Due to the recent discovery of color centers in III-nitride materials, there is renewed interest in the nitrides as a platform for quantum optics. \\

GaN is arguably the most commercially important wide-bandgap semiconductor due to its prevalence in solid state lighting and high power electronics industries. Such industrial interest has accelerated research into GaN, resulting in considerable expertise in epitaxial deposition of complex structures, mature substrate processing and purity, integration on silicon and an understanding of material properties \cite{Huang2017}. However, it wasn't until recently that color centers in GaN were explored as quantum light sources, with emission energies spanning the visible spectrum and the near infrared\cite{Berhane2017,Zhou2018,Nguyen2019}. The large spread in emission energy, as well as differing spectra, suggests these color centers originate from different complexes. Although previously attributed to nanoscale cubic inclusions in the wurtzite host, a report from Nguyen et al.\cite{Nguyen2019} compared various growth regimes and concluded that the emission was likely due to a point defect or impurity, similar to color centers in diamond and silicon carbide \cite{Aharonovich2016}. Theoretical studies of 2-site color centers, such as a subsitutional nitrogen atom adjacent to a vacancy, predicted emission in the \SIrange[repeatunits=false]{1.5}{2}{\electronvolt} range would be observed from more than a dozen configurations \cite{Li2020}. The physical properties of these color centers have yet to be fully understood. \\

Due to the high refractive index of gallium nitride ($n_1 = 2.35$ at \SI{815}{\nano\meter}) single photon extraction is limited by refraction and total internal reflection (TIR) at the interface between the semiconductor and free space. Therefore, a number of enhancement schemes have been investigated and successfully exploited to enhance the extraction efficiency\cite{Barnes2002,Lodahl2015}. One of the most straight-forward methods is to exploit the geometry and high refractive index of a solid immersion lens (SIL)\cite{Serrels2008,Hadden2010}. A SIL is a truncated sphere, commonly of either a hemispherical or Weierstrass geometry, that provides aberration free imaging at one of two aplanatic points. The hemispherical SIL maps the full numerical aperture (NA) of the imaging system into the high index material without refraction at the high-to-low index interface. In addition, the enhancement is inherently broadband and can be applied after pre-selection of a certain color center, with coarse alignment of position. Here, we report the use of a zirconium dioxide (ZrO\textsubscript{2}) hemispherical SIL, which is almost index matched to GaN ($n_{SIL} = 2.13$ at \SI{815}{\nano\meter}), to show enhanced imaging and photon collection from a color center in GaN. \\

\begin{figure*}
    \centering
    \includegraphics{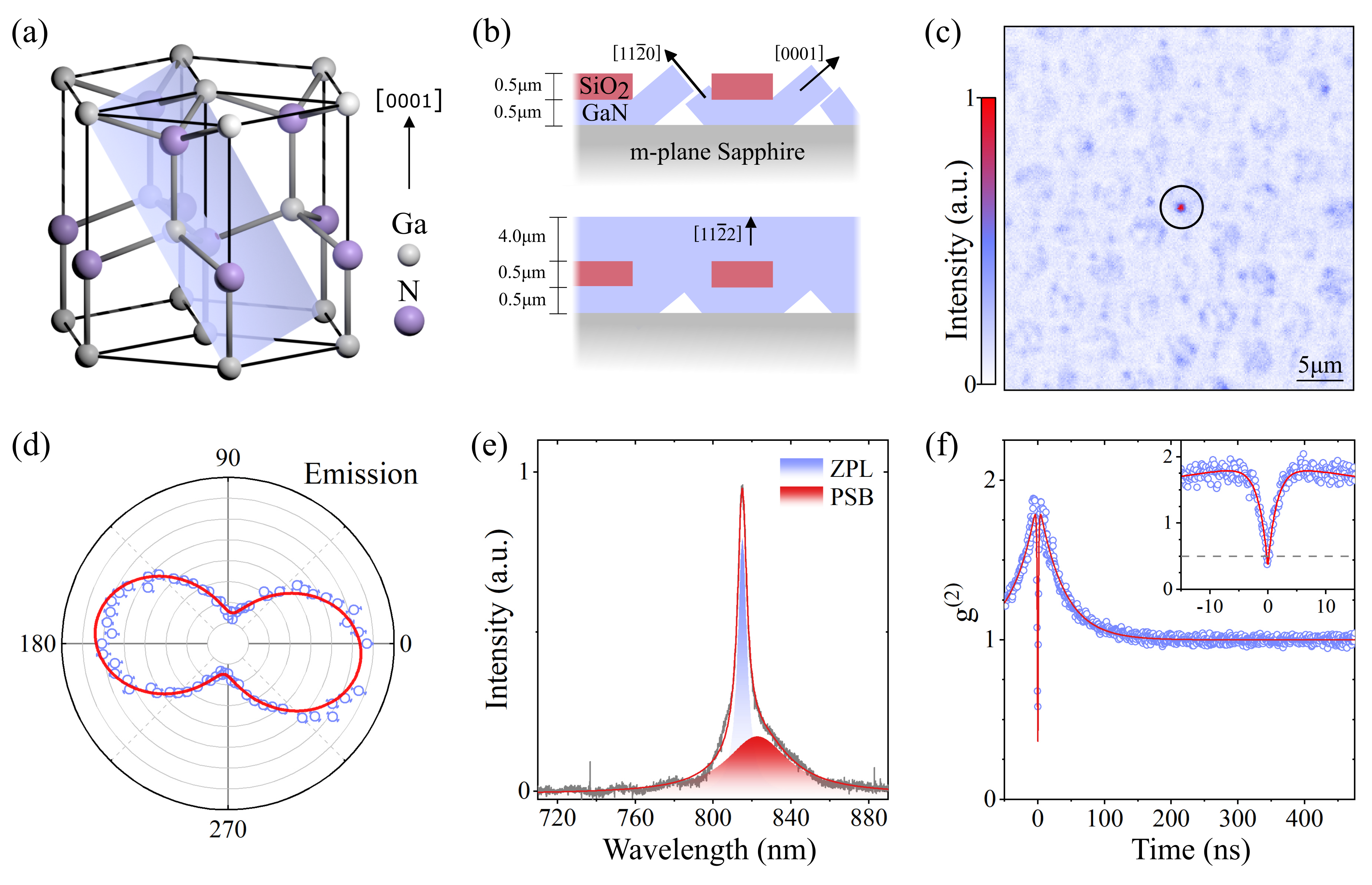}
    \caption{Structural and spectral properties of a color center in semi-polar GaN. a) Crystal structure of Wurtzite GaN. b) Illustration of the growth scheme used to form a semi-polar [$11\bar{2}2$] surface facet. c) Confocal scan map of the sample, with localised emission from the color center seen in the center of the map. d) Polarisation resolved measurement of the emission from the defect. e) Spectral analysis of the color center shown in c). A clear ZPL can be distinguished (purple) broadened by a phonon side band (red). f) Second-order correlation measurement demonstrating clear bunching and anti-bunching.}
    \label{fig:sample}
\end{figure*}

The sample under investigation in this report is a semi-polar gallium nitride on sapphire substrate. The crystal structure of c-plane GaN is shown in Fig.\ref{fig:sample}(a), where the shaded plane represents the $[11\bar{2}2]$ plane which forms the sample surface. The sample structure, illustrated in Fig.\ref{fig:sample}(b), is as follows; Half a micron of m-plane GaN is grown on a m-plane sapphire substrate via metal-organic chemical vapour deposition. A silicon dioxide (SiO\textsubscript{2}) layer is deposited and subsequently patterned, forming circular islands in the m-plane GaN. The sample is then overgrown with GaN. Growth conditions are chosen so the [$0001$] plane overgrows the [$11\bar{2}0$] plane, which results in the semi-polar [$11\bar{2}2$] surface.  \\

A room temperature scanned confocal fluorescence map of the sample, taken with an excitation wavelength $\lambda_{exc} = \SI{532}{\nano\meter}$, is shown in Fig.\ref{fig:sample}(c). An optical long pass filter along with the absorption of the silicon avalanche photodiode (APD) detector used limits the detection window of the collection path of the microscope to between \SI{800}{} and \SI{1000}{\nano\meter} respectively. The confocal scan map reveals the presence of point-like color centers, where the size of the color centers in the image is consistent with a point-like emitter convolved with the diffraction-limited point-spread function of the microscope. This sample displays a much lower density of color centers than c-plane GaN samples that have previously been reported \cite{Nguyen2019}, assisting in the isolation of a single color center. Strongly localised emission below the bandgap suggests the emission originates from one color center. In addition to the color center, the periodic arrangement of embedded disks can be seen in the scan image, due to laser scattering from the SiO\textsubscript{2} disks and/or fluorescence from other photo-active defects. \\

Point-like color centers can be observed in multiple locations in the scan map, both near to and away from the SiO\textsubscript{2} disks, suggesting the origin of the emission lies within the semiconductor and not the SiO\textsubscript{2}. There is often emission between the disks, which display photo-induced bleaching and no antibunching: we do not study these areas further and believe the origin of this emission is different to that of the point like emitters in this report. \\

We focus in this paper on one representative emitter at the center of the scan which is bright as a result of its nanosecond timescale radiative lifetime, and is easy to relocate. The highlighted emitter is used for all measurements presented in this paper. Repeated measurements of the color center over several months have shown no change in its properties. \\

A polarization resolved measurement, shown in Fig.\ref{fig:sample}(d), where a thin film polarizer in the collection path of the microscope was rotated in the plane of the sample, shows dipole-like emission from the color center. We hypothesise that the non-perfect extinction of the polarisation suggests an out of plane component arising from a single dipole aligned at an angle of $\phi=\SI{29.2\pm0.7}{\degree}$ to the sample plane. We note that the non-perfect extinction may also be a result of multiple emission dipoles like what is seen for the NV color center\cite{Dolan2014}. The azimuthal angle $\phi$ is determined from the fit using Eq.\ref{eq:polar} \cite{Barnes2002}, where the emitted intensity $I$ is a function of the in-plane dipole angle $\theta$ and $\phi$, with $n$ as a normalisation factor.
\begin{equation}
    I(\theta,\phi) \propto n[1-sin^2(\theta)cos^2(\phi)]
    \label{eq:polar}
\end{equation}

A spectral measurement of the color center, taken under \SI{532}{\nano\meter} excitation, is presented in Fig.\ref{fig:sample}(e). The spectral measurement is fit using two peaks. A clear zero-phonon line (ZPL) centred at \SI{815}{\nano\meter} can be identified, which arises from the optical excited-to-ground state transition of the defect without coupling to high energy phonons. The symmetric broadening around the ZPL is consistent with coupling to low energy acoustic phonons. The linewidth of the ZPL is measured as \SI{5.75\pm0.03}{\nano\meter}. The ZPL is accompanied by a red-shifted phonon side band (PSB). The ZPL and the contributions from phonon assisted transitions overlap in the spectrum at room temperature, consistent with other color centers in the sample. The portion of the intensity situated within the ZPL, including the symmetric broadening due to acoustic phonons, is \SI{39}{\percent}. The overall spectrum is significantly narrower than that observed from the NV center in diamond \cite{Hadden2010} and AlN \cite{Bishop2020} color centers, in which the phonon sideband dominates the spectrum with a room temperature broadening greater than \SI{100}{\nano\meter}.\\

The photon statistics of the light are measured in the second-order correlation measurement presented in Fig.\ref{fig:sample}(f) taken near saturation power for the emitter. Strong anti-bunching occurs on a \SI{0.90\pm0.02}{\nano\second} timescale while bunching can also be observed on a \SI{37.2\pm0.9}{\nano\second} timescale. The insert shows the antibunching in more detail. A three-level energy model is used to fit the data, where the proposed third level represents a metastable or "shelving" state, consistent with other works on single room temperature color centers \cite{Kurtsiefer2000,Berhane2017,Bishop2020}. The finite value of the correlation function at zero delay can be accounted for due to imperfect filtering of laser scatter and sapphire fluorescence beyond \SI{750}{\nano\meter}. We do not correct for background fluorescence in the measurement. This poissonian background light could be reduced by narrowing the range of the spectral filtering. \\

\begin{figure}
    \centering
    \includegraphics{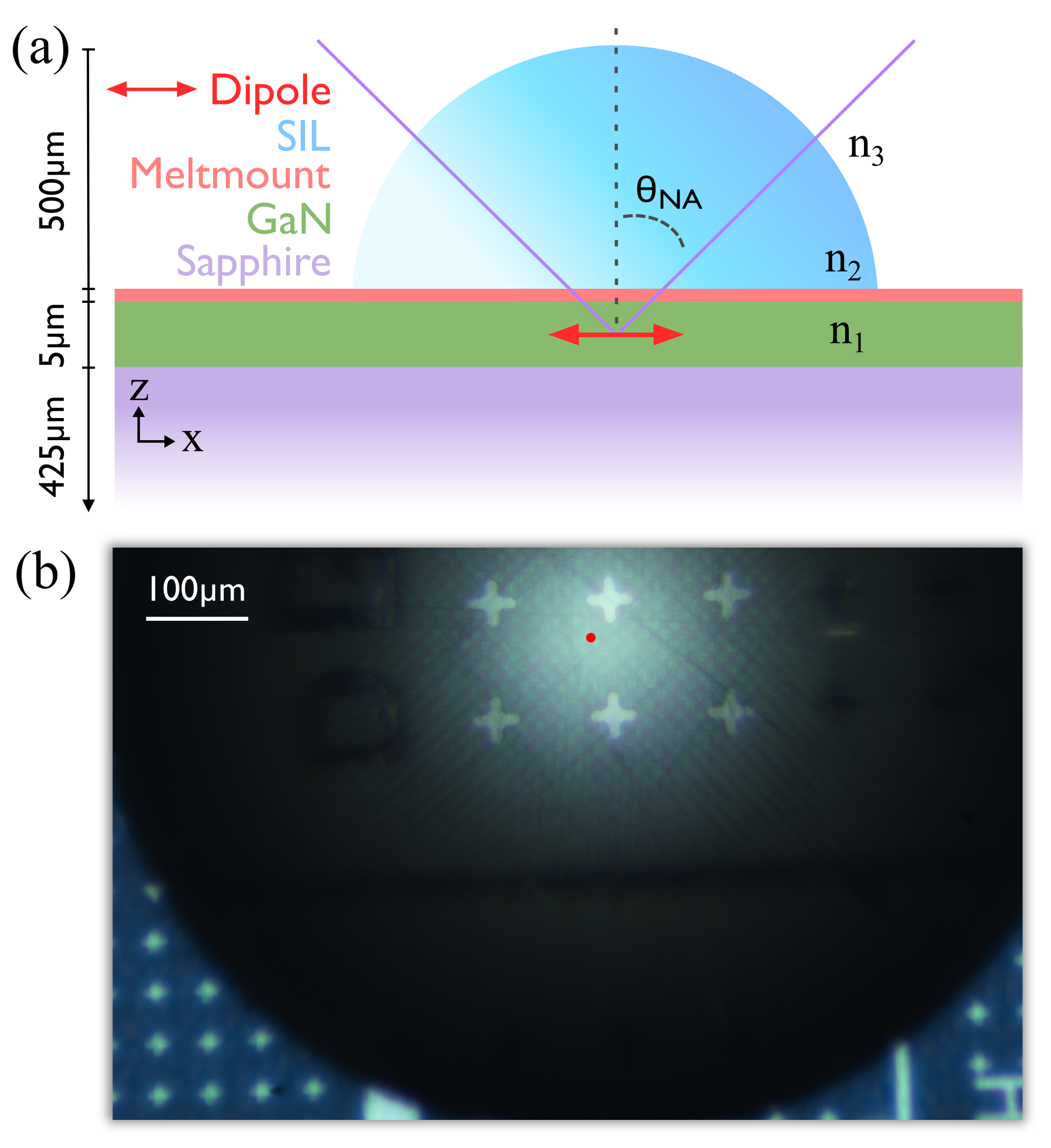}
    \caption{Hemispherical solid immersion lens. a) Cartoon of the geometric properties of the hemispherical SIL. b) Optical image of the sample through the SIL after transfer. The optical magnification caused by the SIL can be seen by comparing the size and separation of the crosses near the center of the SIL with those off the SIL. The location of the emitter is marked with a red dot.}
    \label{fig:SIL}
\end{figure}

\begin{figure}
    \centering
    \includegraphics{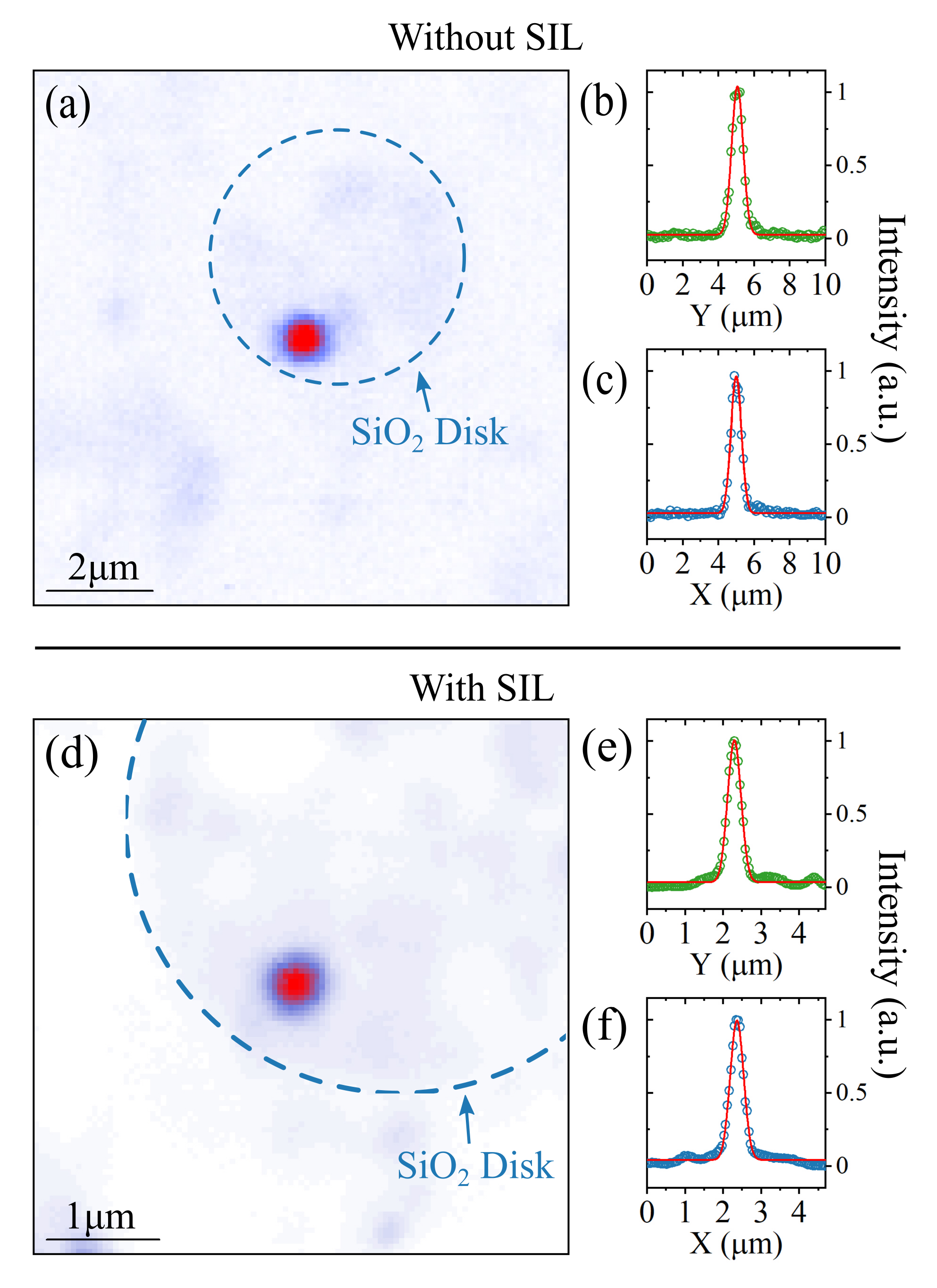}
    \caption{Enhanced imaging of a color center with a solid immersion lens. a) and d) Confocal scan maps around the defect without and with a SIL respectively. The dashed blue lines illustrate the embedded SiO\textsubscript{2} disks. b) and c) X and Y cross-sections of the scan map in a). e) and f) Confocal cross-sections in X and Y for the scan map of the same defect with the SIL. All cross-sections are fit with Gaussian functions.}
    \label{fig:imaging}
\end{figure}

To enhance the light collection from the emitter studied in Fig. \ref{fig:sample}, we deterministically position a \SI{1}{\milli\meter} diameter hemispherical SIL directly on top of the emitter presented in Fig.\ref{fig:sample} using a micromanipulator. This sample was patterned with a regular array of labelled crosses in a thin metallic layer, to allow the position to be accurately determined during placement, with an accuracy of $\pm \SI{10}{\micro\meter}$. A high refractive index mounting media (Cargille Meltmount\texttrademark,  $ n=1.704$) fills the gap between the sample and SIL. \\

The increased NA within the semiconductor allows smaller features to be resolved in the lateral direction \cite{Serrels2008}, where the Rayleigh limit is proportional to 1/NA. Similarly, the increased NA reduces the Rayleigh range of the focused beam, allowing smaller features to be resolved in the axial direction. In addition, an index-matched SIL increases the magnification by a factor of $n_{SIL}$ in the lateral direction. The magnification within the SIL can be seen in the optical image in Fig.\ref{fig:SIL}(b), by comparing the apparent sizes of the patterned crosses visible under the SIL with those in the bottom left hand side of the image. \\

For an aberration free confocal imaging system, the diffraction limited resolution is given by the Rayleigh criterion;
\begin{equation}
    \Theta_{FWHM} = 0.5014 \frac{\lambda_c}{NA}
\end{equation}
where $\lambda_c$ is the collection wavelength, in this case \SI{815}{\nano\meter}. The increased resolution that can be achieved with a SIL, over a sample with a flat surface, is a result of the increased numerical aperture of the light coupled into the semiconductor, $NA_{SIL} = n_2 \times NA_{Imaging}$.\\

The enhanced resolution is illustrated in Fig.\ref{fig:imaging} which includes two scans over the same angular range, with and without the SIL. The embedded SiO\textsubscript{2} disks can be seen in both scan maps and are highlighted with the dashed blue lines. Clear magnification of the image can be seen evidenced by the increased disk size, by a factor of $n_{SIL}$. The magnification is also made apparent in the optical image shown in Fig.\ref{fig:SIL}(b), in which one can compare the angular dimensions of identical lithographic features beneath the SIL to those on the flat surface. In the two scans in Fig.\ref{fig:imaging} the increase in magnification of the imaging system with the SIL cancels the increased resolution, resulting the emitter having the same angular size. X and Y slices across the emitters are fitted using a Gaussian function and the FWHM compared to the expected spot size is presented in Table.\ref{tab:resolution}.  The uneven sample surface and the chromatic behaviour of the objective prevents our imaging system reaching the resolution limit, but nevertheless, the data shows a clear enhancement in resolution with the SIL. \\

\begin{table}
\begin{ruledtabular}
\caption{\label{tab:resolution} Imaging resolution with and without the SIL as compared to the Rayleigh criterion.}
\begin{tabular}{lccc}
& X (nm) & Y (nm) & Rayleigh Criterion (nm)\\
\hline
No SIL & $693\pm9$ & $755\pm9$ & $544$\\
With SIL & $416\pm6$ & $417\pm6$ & $255$\\
\end{tabular}
\end{ruledtabular}
\end{table}

The extraction of light from a high refractive index material is limited due to refraction at the sample surface. Even with a high NA lens, there is a limited light collection efficiency due to total internal reflection at the semiconductor-to-air interface, $\theta = \arcsin{(n_3/n_1)} = \SI{26}{\degree}$. For the case where a ZrO\textsubscript{2} hemispherical SIL is placed directly on top of the semiconductor, the numerical aperture within the sample is close to that of the lens as all light exits the SIL normal to the SIL-to-air interface, therefore increasing the number of photons collected.\\

\begin{figure}
    \centering
    \includegraphics{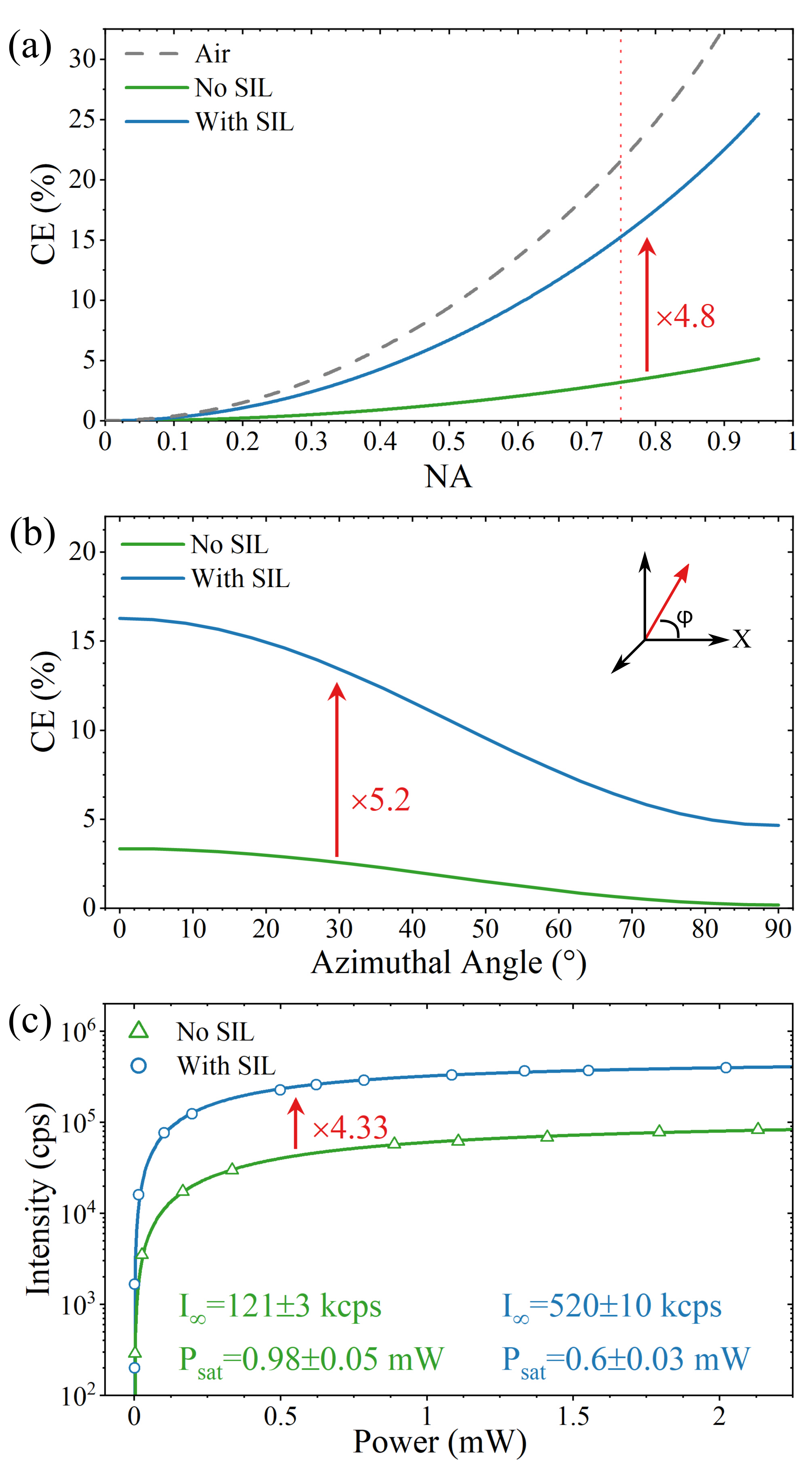}
    \caption{Collection efficiency enhancement with a hemispherical SIL. a) Calculated collection efficiency enhancement for an in-plane dipole as a function of the NA, calculated using Eq.\ref{eq:CE}, for a dipole in air (grey), in GaN (green) and in GaN with a ZrO\textsubscript{2} SIL (blue). b) Numerically determined collection efficiency as a function of the azimuthal angle, determined using finite different time domain simulations. c) The measured collection efficiency enhancement of the SIL, taken using power dependent measurements on the same emitter with and without the SIL. A collection efficiency enhancement of $\times\SI{4.3\pm0.1}{}$ is achieved.}
    \label{fig:collection}
\end{figure}

For a dipole source orientated in-plane it is possible to determine the collection efficiency into a lens of numerical aperture NA=0.75 using an analytical expression, as presented by Barnes et.al \cite{Barnes2002};
\begin{equation}
    \eta = \frac{1}{32}\Bigg[15\bigg(1-\sqrt{1-sin(\theta)^2}\bigg)+\bigg(1-cos(3\theta)\bigg)\Bigg]T_{a}
    \label{eq:CE}
\end{equation}
where, for a color center in bulk GaN;
\begin{eqnarray*}
    \theta=\arcsin(\frac{NA}{n_1}) 
    \quad \text{and} \quad
    T_a = \frac{4n_1n_3}{(n_1+n_3)^2}\\
    \\\quad \therefore \quad 
    \eta_{bulk} \approx \SI{3.2}{\percent}
    \label{eq:fresh}
\end{eqnarray*}
and with the SIL;
\begin{eqnarray*}
    \theta=\arcsin(\frac{n_{SIL}NA}{n_1n_3}) 
    \quad \text{and} \quad
    T_a = \frac{16n_1n_{SIL}^2n_3}{(n_1+n_{SIL})^2(n_{SIL}+n_3)^2}\\
    \\\quad \therefore \quad 
    \eta_{SIL} \approx \SI{15.3}{\percent}
\end{eqnarray*}
$T_a$ is an approximation of the transmission due to Fresnel reflections at the GaN-to-air and SIL-to-air interfaces respectively. Therefore, we predict an improvement in the collection efficiency with the SIL equal to $\times 4.8$. The collection efficiency as a function of the numerical aperture for a dipole in air (grey, dashed), in GaN (green) and with the SIL (blue) is illustrated in Fig.\ref{fig:collection}(a). The significant enhancement of the collection efficiency with the SIL is apparent as the blue plot approaches that of the ideal case of a dipole in air.\\

To determine the expected collection efficiency enhancement for a dipole with a finite azimuthal angle we turn to Finite Difference Time Domain (FDTD) simulations using commercial software (Lumerical). The GaN epilayer is considered to be a homogeneous dielectric layer without the SiO\textsubscript{2} disks and no absorption. The electric field profile is measured in three dimensions around the dipole. The near-field electric field profile above the surface in the XY plane is projected into the far-field and integrated across the half angle of the collection optics (NA=0.75). Due to the geometry of the SIL, the near field electric field profile is collected in the ZrO\textsubscript{2} dielectric and projected into the far field. The reflection at the SIL-to-air interface is accounted for using $T_a$ in Eq.\ref{eq:CE}. Due to the unknown thickness of the adhesion layer used in the experiment, we do not account for the Meltmount in the simulation. The results therefore present an upper bound for the collection efficiency enhancement of an out of plane emitter.\\

As illustrated in Fig.\ref{fig:collection}(b), where the dipole is in the plane of the sample ($\phi=0$), the CE agrees with the analytically determined value in Eq.\ref{eq:CE} with and without the SIL, with a CE of \SI{16.3}{} and \SI{3.4}{\percent} respectively. The expected enhancement at the determined azimuthal angle $\phi=\SI{29.2\pm0.7}{\degree}$ is slightly increased from the in-plane case, with an enhancement of $\times5.2$. \\

The measured collection efficiency enhancement is shown in (c), where the power dependent intensity of the color center is presented with and without the SIL. The data is fit with a saturation function, $I(P) = I_{\infty}\frac{P}{P+P_{sat}}$, where $I_{\infty}$ is the count rate at infinite power and $P_{sat}$ is the power required to saturate the emitter. $I_{\infty}$ is used to quantify the enhancement, where $I_{\infty}$ with and without the SIL is \SI{520\pm10}{kcps} and \SI{121\pm3}{kcps} respectively, an improvement equal to $\times\SI{4.3\pm0.1}{}$. The discrepancy with the predicted value may be a result of the buried SiO\textsubscript{2} disks and rough sample/SIL surface. \\

We investigated the properties of color centers in semi-polar GaN, demonstrating room temperature quantum light in the near infrared from a color center with a dipole-like emission pattern. Enhanced imaging of a single emitter in a semi-polar GaN sample was achieved by spatially aligning a SIL on top of a color center. The use of the hemispherical SIL also increased the photon collection efficiency by a factor of $\SI{4.3\pm0.1}{}$, such that the efficiency approaches that of a dipole in free space. The combination of existing expertise in GaN processing and epitaxy, with enhanced photon extraction enabled by this easy-to-implement SIL, can lead to efficient off-the-shelf quantum light sources operating without cryogenic cooling, which will be highly beneficial for future quantum technologies. \\

\section*{acknowledgement}
The authors acknowledge financial support provided by the Sêr Cymru National Research Network in Advanced Engineering and Materials, the H2020 Marie Curie ITN project LasIonDef (GA n.956387), EPSRC grant EP/T017813/1 and feasibility study "ACES" (funded via EP/P006973/1).

\section*{Data Availability Statement}

The data that support the findings of this study are openly available in the Cardiff University Research Portal at http://doi.org/[doi], reference number[reference number]..\\

\bibliography{ms.bib}
\end{document}